%% file: paper_v4_costas.tex
\documentclass[12pt]{article}  
\usepackage{times} 
\usepackage{fleqn} 
\usepackage{amssymb,bm} 
\usepackage{epsfig,graphicx,color}  
\usepackage{import} 

\newcommand\be{\begin{eqnarray}}
\newcommand\ee{\end{eqnarray}}

\begin{document}

\addtolength{\baselineskip}{0.20\baselineskip}

\vspace{48pt}

\centerline{\Huge The Logarithmic Triviality of Compact QED}
\medskip
\centerline{\Huge Coupled to a Four-Fermi Interaction }

\vspace{28pt}

\centerline{\bf John B. Kogut$^a$ and
Costas G. Strouthos$^b$}
                                                                                                                                      
\vspace{15pt}
                                                                                                                                      
\centerline{$^a$ {\sl Physics Department, University of Illinois at Urbana-Champaign,}}
\centerline{\sl Urbana, IL 61801-30.}
\smallskip
\centerline{$^b$ {\sl Department of Physics, University of Cyprus,}}
\centerline{\sl CY-1678 Nicosia, Cyprus.}
\vspace{24pt}
                                                                                                                                      
                                                                                                                                      
\centerline{{\bf Abstract}}
                                                                                                                                      
\noindent
{\narrower
This is the completion of an exploratory study of
Compact lattice Quantum Electrodynamics with a weak four-fermi interaction 
and 
four species of massless fermions. In this formulation of  Quantum Electrodynamics  
massless fermions can be simulated directly and
Finite Size Scaling analyses can be performed at the theory's chiral symmetry
breaking critical point.
High statistics simulations on lattices ranging from
$8^4$ to $24^4$ yield the equation of state, critical indices, scaling functions and cumulants.
The measurements are well fit with the orthodox
hypothesis that the theory is logarithmically trivial and its continuum limit 
suffers from Landau's zero charge problem.
}


\vfill
\newpage

\section{Introduction}

This is the final paper in a series of simulation studies searching for a
nontrivial ultra-violet fixed point in abelian gauge theories. Previous studies
have provided simulation evidence that scalar QED \cite{baig} and noncompact lattice QED with four
species of massless fermions \cite{PLB}
are logarithmically trivial.

This work employs the $\chi$QED formulation of the model in which
a weak four-fermi interaction is added to the standard action. This affords us
two advantages over standard methods: 1. we can simulate massless fermions directly 
on the lattice and see how massless
fermion charge screening affects the dynamics, and 2. the four-fermi interaction separates
the chiral transition of the model from its confinement/deconfinement
transition which is a first order transition and is controlled by monopole condensation.

The $\chi$QED action contains two couplings, $\beta=1/e^2$ where $e$ is the usual 
electrodynamic charge, and the four-fermi coupling, $G=1/\lambda$. The model's phase diagram
contains two separate lines of transitions, one describing monopole condensation 
which is first order and a line of second order chiral transitions.
We will show that the line of second order chiral transitions
describes a logarithmically trivial continuum limit, presumably identical to the continuum
limit of the noncompact $\chi$QED lattice model studied earlier \cite{PLB}.

This will be accomplished by exploiting the fact that Finite Size Scaling (FSS) applies
simply to $\chi$QED because it is formulated without a bare fermion mass. In the conventional
lattice action the bare fermion mass explicitly breaks chiral symmetry and introduces scale breaking.
A nonzero fermion mass
is needed so that the standard algorithm converges. By contrast, $\chi$QED allows us to
study the physics of the critical point by
doing simulations in its immediate vicinity and extracting information from FSS arguments
without ever needing uncontrolled extrapolations to the chiral limit.
One of the purposes of this paper is the illustration of FSS methods
for lattice gauge theories with massless fermions. This represents new territory for lattice gauge theory.

Of course, this work is not without its disappointments. We cannot simulate the model in a range of parameters
where monopoles are relevant degrees of freedom in the theory's continuum limit.
For this action, the line of monopole condensation
transitions is distinct from that of the chiral transitions and monopole condensation appears
to be first order. It has been argued elsewhere \cite{KH} that
a second order transition where there
is both chiral symmetry breaking and monopole condensation would be a natural scenario for a nontrivial
form of continuum QED because screening effects coming from fermions could balance anti-screening
effects coming from monopoles. A future formulation of lattice monopoles and fermions will have to be developed
to see if this possibility could work out.

Since this paper is a continuation in a series we refer the reader to reference \cite{KC} for additional background
and formulas. Here our emphasis is on the new simulations, FSS methods and
results.

The paper begins by laying out the lattice theory's two dimensional phase diagram and presenting some simulations
which pin down its qualitative features. Then some highly accurate
$16^4$ and $24^4$ simulations in the broken symmetry phase
are presented and analyzed. This is the approach used in past studies which showed that the noncompact model is
logarithmically trivial \cite{PLB}. The compact model should also be logarithmically trivial because the chiral
transition occurs in a region of the model's phase diagram where monopoles are not critical. Under these conditions
the differences between the compact and noncompact models should become irrelevant and the models should
have the same continuum limit \cite{MBK}. This conventional idea was not well supported by our first 
simulation study \cite{KC}
of the model which motivated us to do a more thorough job which we report here.
The conclusions presented here are based on over ten times the statistics of earlier work. High statistics 
allowed the simulation program to reach
thermodynamic equilibrium and produce more accurate ensembles of configurations which reduced the
error bars substantially. The $16^4$ and $24^4$ data in the broken symmetry phase can be
fit with both power laws whose critical indices deviate from mean field theory, as reported in \cite{KC},
\emph{or} log-improved mean field theory. The log-improved mean field theory fits are somewhat
better than the power law fits, but a conclusive result eludes us if we just use this subset
of our simulations. However, the combination of the broken phase results together with new simulations,
analyzed with FSS methods on lattices ranging from $8^4$ to $16^4$, in the immediate vicinity
of the model's critical point
lead to stronger results: they favor the logarithmically trivial fermionic field theory scenario.
This is the new conclusion which will be presented below.

This paper is organized as follows: In the next section we briefly review the formulation
of the lattice action and define the parameters we use in our fits and simulations. In the third section
we sketch the phase diagram and present a few simulations which led to it. Then we examine
several points in the phase diagram along the line of chiral transitions and show that the
data for the chiral order parameter in the broken symmetry phase is compatible with log-improved mean field theory.
Next we review the relevant features of FSS and present several sections
of analysis of high statistics data sets using lattices ranging from $8^4$ through $16^4$. 
This is the most decisive analysis in this study.
Its success depended crucially on
the use of the $\chi$QED action and statistically large data sets which are several orders of magnitude larger
than those used in typical lattice simulations of QCD, for example. FSS simulations at and 
near the critical coupling were essential to establishing the logarithmic triviality of this model.

\section{Formulation}

The lattice Action of compact $\chi$QED, where the gauge symmetry is interpreted
as a compact local $U(1)$ symmetry, following Wilson's original proposal \cite{wilson}, reads

\begin{equation}
S  =  \sum_{x,y} \bar\psi(x) (M_{xy} + D_{xy}) \psi(y) +
  \frac {1}{2 G} \sum_{\tilde x} \sigma ^2 (\tilde x) +
  \frac{1}{2 e^2} \sum_{x,\mu,\nu} (1-\cos(F_{\mu\nu}(x))),
\end{equation}

\noindent
where
\begin{eqnarray}
F_{\mu\nu}(x)& = &\theta_{\mu}(x) + \theta_{\nu}(x+\hat{\mu}) +
\theta_{-\mu}(x+\hat{\mu}+\hat{\nu}) + \theta_{-\nu}(x+\hat{\nu}), \\
M_{xy}& = & (m + \frac{1}{16} \sum_{\langle x,\tilde x \rangle} \sigma( \tilde x))
\delta_{xy}, \\
D_{xy} & = & \frac{1}{2} \sum_\mu \eta_{\mu} (x) (
 e^{i\theta_{\mu}(x)} \delta_{x+\hat{\mu}, y}
- e^{-i\theta_{\mu}(y)} \delta_{x-\hat{\mu}, y} ).
\end{eqnarray}

\noindent
The auxiliary scalar field $\sigma$ is defined on the sites of the dual
lattice $\tilde x$ \cite{CER}, and the symbol $\langle x,\tilde x \rangle$ denotes
the set of the 16 lattice sites surrounding the direct site $x$.
The factors $e^{\pm i\theta_\mu}$ are the gauge connections and
$\eta_\mu(x)$ are the staggered phases, the lattice analogs of the
Dirac matrices. $\psi$ is a staggered fermion 
field and  $m$ is the bare fermion mass, which will be set to zero.
Note that the lattice expression for $F_{\mu\nu}$ is the circulation of
the lattice field $\theta_{\mu}$ around a closed plaquette, the gauge field couples to the
fermion field through compact phase factors to guarantee local gauge
invariance and $\cos F_{\mu\nu}$ enters the action to make it compact.

It will often prove convenient
to parametrize results with the inverse of the four-fermi coupling, $\lambda \equiv 1/G$,
and the inverse of the square of the gauge coupling, $\beta \equiv 1/e^2$.

The global $Z_2$ chiral symmetry of the Action reads:

\begin{eqnarray}
\psi(x) & \rightarrow & (-1)^{x1+x2+x3+x4} \psi(x) \\
\bar \psi(x) & \rightarrow & -\bar \psi (x) (-1)^{x1+x2+x3+x4} \\
\sigma & \rightarrow & - \sigma.
\end{eqnarray}
where $(-1)^{x1+x2+x3+x4}$ is the lattice representation of $\gamma_5$.

Interesting limiting cases of the above Action are: (i) the $Z_2$ Nambu$-$Jona-Lasinio
model with no gauge fields, set $e^2$ to zero here, which has a logarithmically trivial
chiral phase transition at nonzero $G$; (ii)
the compact QED model with no four-fermi interactions, whose first order chiral phase transition is 
coincident with its first order monopole condensation transition
near $\beta \equiv 1/e^2 \approx 0.89(1)$ for four flavors \cite{Dag}; and 
(iii) the $G \rightarrow \infty$ limit in which the fermions obtain a dynamical mass
comparable to the reciprocal of the lattice spacing and therefore decouple,
leaving quenched compact QED which has a first order transition at $\beta = 1.011124(1)$ \cite{krauts}.

We refer the reader to earlier papers in this series \cite{KC} for more motivation and details. 
We only emphasize new simulations and analyses here.

\section{Overview of the Phase Diagram.}

\begin{figure}
\begin{center}
\scalebox{.5}{ \input{pqedz2.pstex_t} }
\caption{Phase Diagram of Gauged Compact $U(1)$ Nambu Jona-Lasinio Model}
\label{fig:pqedz2}
\end{center}
\end{figure}
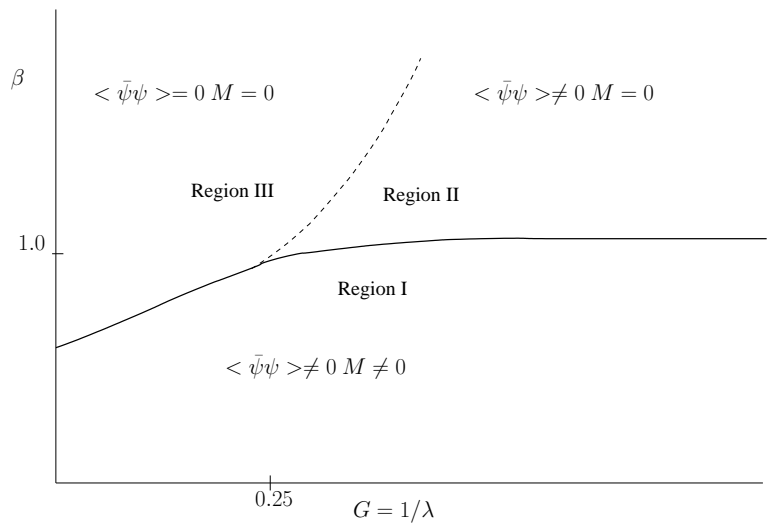

The phase diagram for the model \cite{KC} is shown in  Fig. \ref{fig:pqedz2}.
The monopole concentration $M$ and the chiral condensate, $\langle \bar{\psi} \psi \rangle$, label
the three phases. Region I has chiral symmetry breaking in a condensate of monopoles,
Region II has chiral symmetry breaking in a monopole free vacuum, and
Region III is chirally symmetric
in a monopole free vacuum.
Past simulations suggest that the dashed line consists of second order transitions
and the thick line consists of first order transitions \cite{KC}.

It is interesting to confirm that the dashed line of chiral symmetry breaking transitions
in the upper reaches of Fig. \ref{fig:pqedz2} turns vertical and the four-fermi
coupling alone breaks chiral symmetry at strong coupling, in agreement with
\cite{Looking}. This is shown in Fig.~2 where we confirm that the model breaks
chiral symmetry for $G=1.0$ no matter how small the
gauge coupling $e^2=1/\beta$.
\begin{figure}

\centerline{
\epsfxsize 4.0 in
\epsfysize 3.0 in
\epsfbox{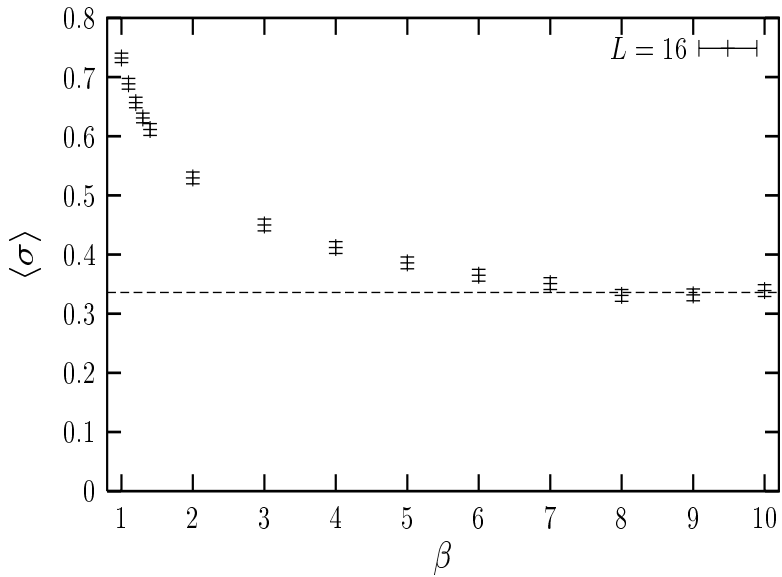}
}
\caption{$\langle \sigma \rangle$ vs. gauge coupling $\beta = 1/e^2$ at fixed four-fermi coupling $G=1.0$.}
\label{fig:largebeta}
\end{figure}

In reference \cite{KC} we concentrated on the vertical line $G=1/1.4$ and showed that the line of
first order monopole concentration transitions is clearly separate from the line of chiral symmetry
transitions. We felt that it was important to simulate along another vertical line in the phase diagram
to verify this result and to see how sensitive the characteristics of the transitions are to the bare
couplings. There are several questions we need to address. They include: (i) Does the order of the transition(s)
change along the lines in the phase diagram? (ii) Do the critical indices change along the lines? There are
related models where this is known to happen, such as in the quenched noncompact gauged Nambu-Jona Lasinio
model \cite{Love}. It is unknown if such phenomena can occur in unquenched abelian models where fermion
screening leads to the zero charge problem in perturbation theory and produces only perturbatively trivial
models.

In Fig.~3 we show the monopole concentration as the gauge coupling $\beta$ passes through
the phase boundary at $G=0.50$ between regions I and II. The monopole concentration appears to have a discontinuous
jump just below $\beta=0.935$.

\begin{figure}
\centerline{
\epsfxsize 4.0 in
\epsfysize 3.0 in
\epsfbox{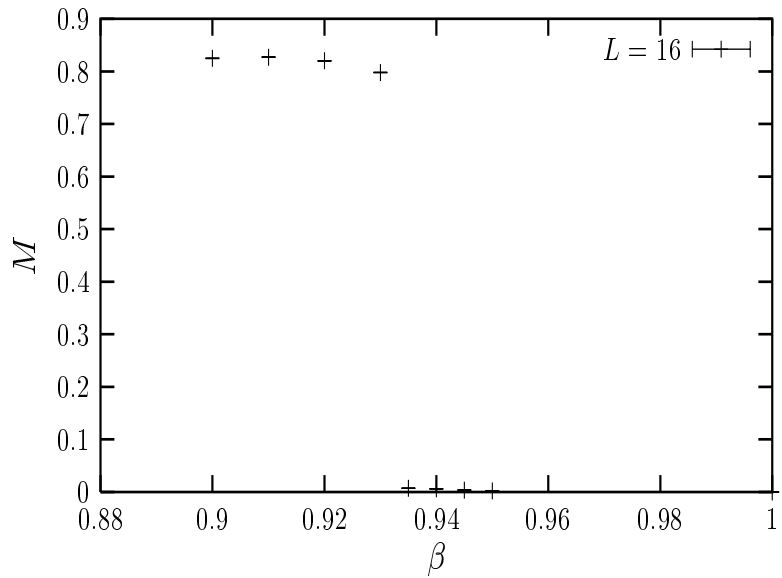}
}
\caption{Monopole concentration vs. gauge coupling $\beta$ at fixed four-fermi coupling $G=0.50$.}
\label{fig:pqed16bs2mon}
\end{figure}

Measurements of $\langle \sigma \rangle$ in Fig.~4
also show the first order transition near $\beta=0.935$ and indicate that 
region II (see Fig.~\ref{fig:pqedz2}) extends from this point to approximately $\beta \approx 0.96$ 
along the vertical line at $G=0.50$.

\begin{figure}

\centerline{
\epsfxsize 4.0 in
\epsfysize 3.0 in
\epsfbox{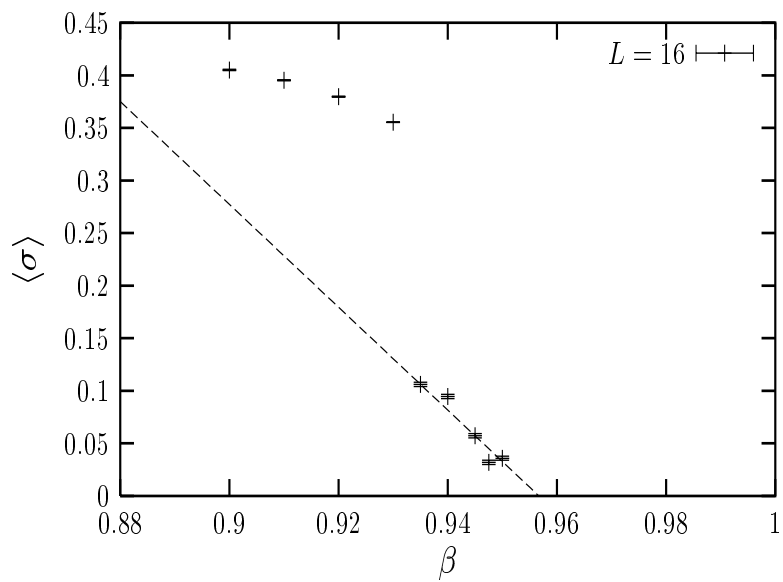}
}
\caption{$\langle \sigma \rangle$ vs. gauge coupling $\beta$ at fixed four-fermi coupling $G=0.50$.}
\label{fig:pqed16bs2sig}
\end{figure}

The simulations reported in \cite{KC} along the vertical line $G=1/1.4$
found a continuous chiral transition at $\beta=1.393(1)$. It is interesting to check the consistency
of this prediction with simulations approaching the critical point from a different direction 
in the phase diagram. Therefore,
we simulated the model on $16^4$ lattices along the horizontal direction, fixing $\beta=1.393$ and varying
the four-fermi coupling. The results are shown in Fig.~5.

\begin{figure}

\centerline{
\epsfxsize 4.0 in
\epsfysize 3.0 in
\epsfbox{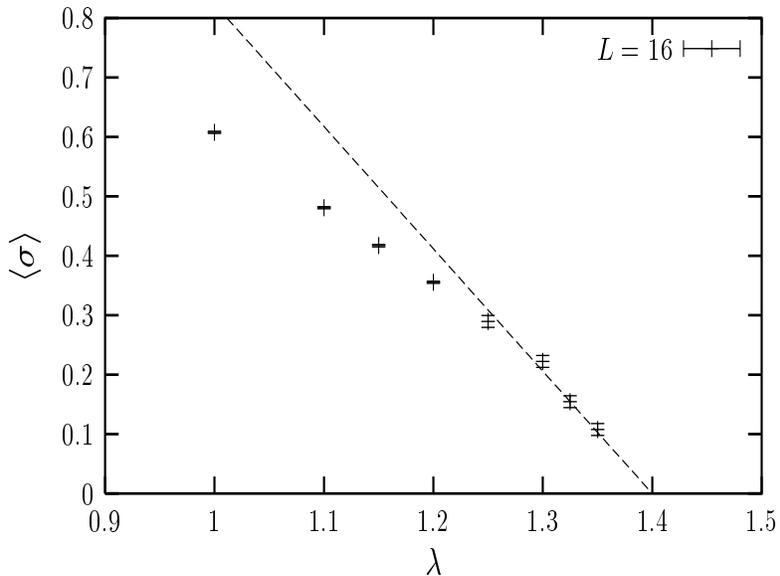}
}
\caption{$\langle \sigma \rangle$ vs. four-fermi coupling $\lambda=1/G$ at fixed gauge coupling $\beta=1.393$.}
\label{fig:pqed16b1393}
\end{figure}

We see that Fig.~5, although not competitive in accuracy with the study at fixed four-fermi coupling and
variable gauge coupling due to an apparently narrow scaling region, is indeed compatible with the
earlier results: the critical point is again predicted to be at $G=1/1.4$ and $\beta=1.393$. The dashed curve
in Fig.~\ref{fig:pqed16b1393} is just meant to guide the eye to the horizontal line.

\section{High Statistics $16^4$ Simulations along the Vertical Line $\lambda=1.4$}

\begin{figure}

\centerline{
\epsfxsize 4.0 in
\epsfysize 3.0 in
\epsfbox{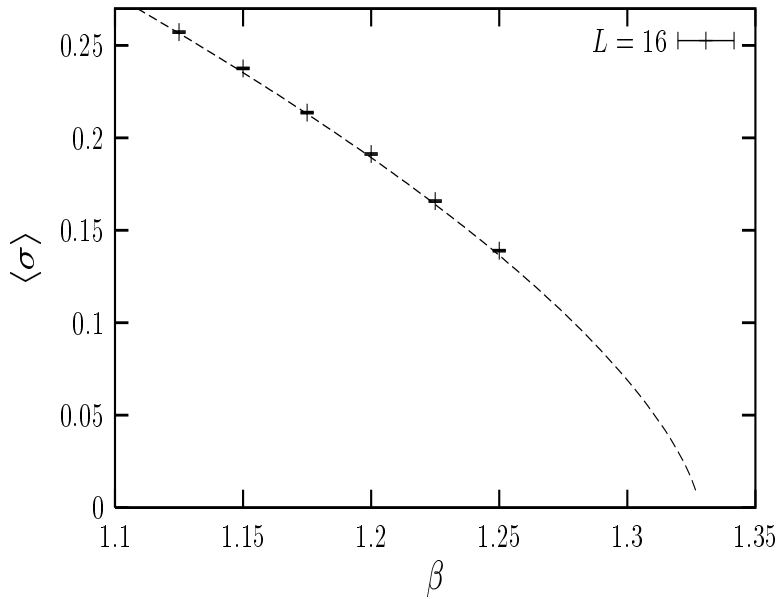}
}
\caption{$\langle \sigma \rangle$ vs $\beta$ at four-fermi coupling $\lambda=1.4$.
The fit has the critical index $\beta_{mag} = 0.65(4)$.}
\label{fig:sig16power}
\end{figure}

The real focus of this series of simulations is to determine the quantitative nature of the line of chiral
transitions in the phase diagram. We, therefore, developed a very fast, parallel version of the simulation
code and made high statistics runs along the vertical line $\lambda=1.4$, complementing the exploratory
runs reported earlier \cite{KC}. In fact, between $10 \times 10^6$ and $20 \times 10^6$ sweeps 
of the Hybrid Molecular
Dynamics algorithm were done at each of fifteen couplings $\beta$ ranging from $\beta = 1.393$ to
$\beta = 1.125$. This is more than an order of magnitude greater statistics than those previously reported and
represent between $100,000$ and $200,000$ trajectories of the Hybrid Molecular Dynamics
algorithm \cite{HMD} at each coupling (the Monte Carlo
time interval was chosen to be $dt = 0.01$ to keep systematic errors negligible in the molecular dynmics steps).
The data on the broken chiral symmetry side of the transition is plotted in Fig.~\ref{fig:sig16power} 
and a power-law
fit is included in the figure. (The data on the other side of the transition will be used in a FSS analysis below.)
The power-law fit, $\langle \sigma \rangle = A(\beta_c-\beta)^{\beta_{mag}}$, is acceptable, 
$\chi^2/DOF = 1.7$ with a critical
point $\beta_c = 1.33(1)$ and a critical magnetization exponent $\beta_{mag}=0.65(10)$.

It is interesting that the critical index is larger than the mean field value of $1/2$, but the deviation from
mean field theory is less than reported in our earlier, lower statistics, exploratory work \cite{KC}. 
The high statistics of this run are making a difference in the results. We recall from other works using
$U(1)$ gauge fields and the simplest Wilson action, that very long relaxation times
are noted \cite{krauts}. Our time
correlation analysis suggests that the $20$ million sweeps used here suffice and the error bars in the
figure conservatively account for the correlations in Monte Carlo time. It is interesting that
simulations of the noncompact model \cite{PLB} did \emph{not} require such enormous statistics to achieve
equally accurate results.

\begin{figure}

\centerline{
\epsfxsize 4.0 in
\epsfysize 3.0 in
\epsfbox{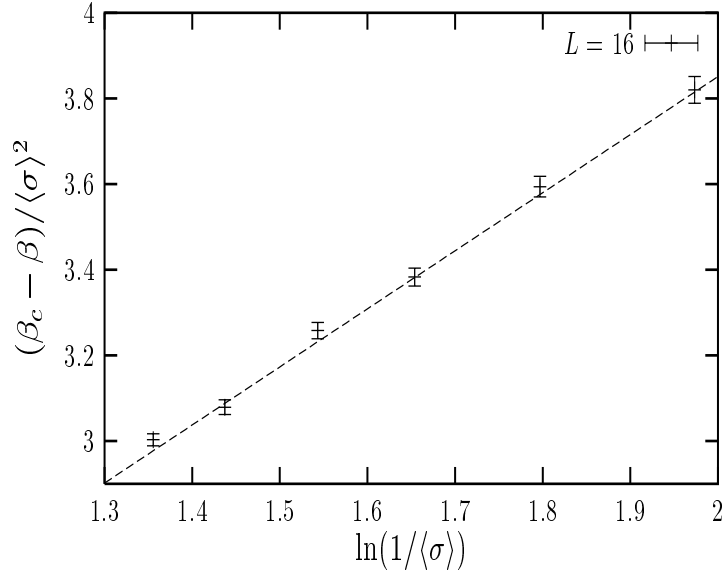}
}
\caption{$(\beta_c-\beta)/\langle \sigma \rangle^2$ vs $\ln (1/\langle \sigma \rangle)$ 
at Four Fermi coupling $\lambda=1.4$.}
\label{fig:sig16}
\end{figure}

Following our analysis of the noncompact model, however, it is interesting to attempt to fit the
same data with the hypothesis of logarithmic triviality. As discussed in previous work, including reference
\cite{Looking}, the logarithms of triviality effect the scaling laws and equation of state differently
for fermionic theories than for bosonic theories. In particular, in references~\cite{PLB,Looking}, the
leading order equation of state had the form 
$\beta_c-\beta=A \langle \sigma \rangle^2 (\ln (1/\langle \sigma \rangle) +B)$. Fits of this
form accomodate the data of Fig. \ref{fig:sig16power} very well, in fact, just like in the noncompact model.
Fitting routines predict the parameter $B=0.84(8)$, 
$A=7.3(9)$ and $\beta_c=1.33(1)$ with $\chi^2/DOF=0.66$.
To emphasize the need for the logarithm here, we plot the quantity $(\beta_c-\beta)/\langle \sigma \rangle^2$ against
$\ln (1/\langle \sigma \rangle)$ in Fig.~\ref{fig:sig16}, and show the fit. 
This makes the point that the simulation results
are compatible with logarithmic triviality. In fact, they appear to rule out
bosonic triviality fits which would have the
logarithm of triviality in the denominator of the equation, 
like $\beta_c-\beta=A \langle \sigma \rangle^2 /(\ln (1/\langle \sigma \rangle) +B)$,
rather than the numerator. Our work, therefore, supports the analytic predictions of
A. Koci$\acute{c}$ \cite{kocic}.
Perhaps, the fact that the $\chi^2/DOF$
from the log-improved mean field relation (which is a three-parameter function) is 
less than half the $\chi^2/DOF$ we get from the fit to the standard power-law 
relation (which is also a three-parameter function) 
is evidence that the data favor the triviality scenario over the interacting field
theory scenario. More compelling and straight-forward evidence for triviality will be
presented in the sections on FSS below.

This result is not above criticism, however. As is clear from the figures, we are not able to simulate
the model very close to the critical point on this lattice size without meeting uncontrollable finite size effects.
For example, simulations closer to $\beta_c$ display tunneling between the $Z_2$ vacua 
through the unbroken phase and make reliable
measurements of the order parameter impossible. We will turn to FSS simulations below to remedy this limitation
and obtain a better estimate for $\beta_c$.

\section{$24^4$ Simulations along the Vertical Line $\lambda=2.0$}

\begin{figure}

\centerline{
\epsfxsize 4.0 in
\epsfysize 3.0 in
\epsfbox{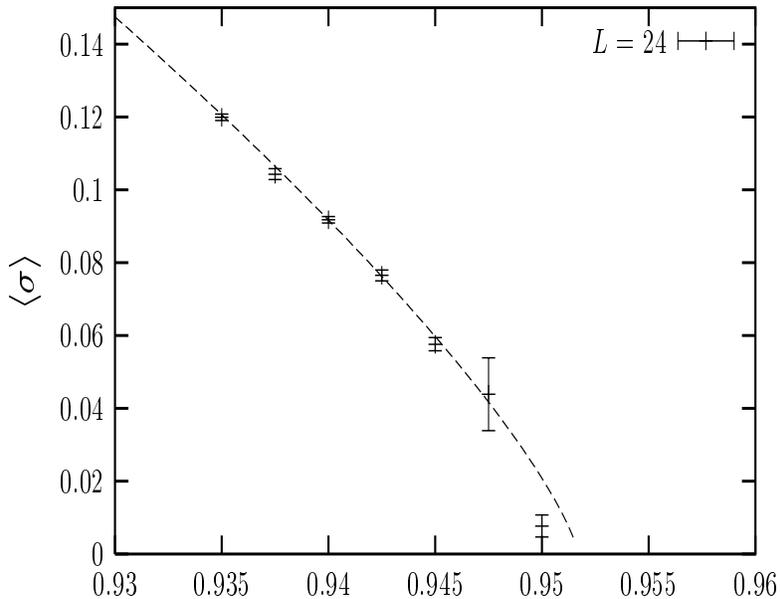}
}
\caption{$\langle \sigma \rangle$ vs $\beta$ at Four Fermi coupling $\lambda=2.0$.
The fit has the critical index $\beta_{mag} = 0.77(10)$.}
\label{fig:pqed24bs2sig}
\end{figure}

Next consider simulations on larger lattices in
the broken symmetry phase. In Fig.~\ref{fig:pqed24bs2sig} we show the raw data from simulations
on $24^4$ lattices at a somewhat weaker four-fermi coupling $G=0.5$. These larger lattices 
allowed us to run simulations closer to the critical point without suffering from large finite size effects.
The algorithm was also somewhat better behaved at weaker four-fermi coupling $G=0.5$.
Of course we were not able
to amass as high statistics in this case: one million sweeps per coupling were accumulated.

\begin{figure}

\centerline{
\epsfxsize 4.0 in
\epsfysize 3.0 in
\epsfbox{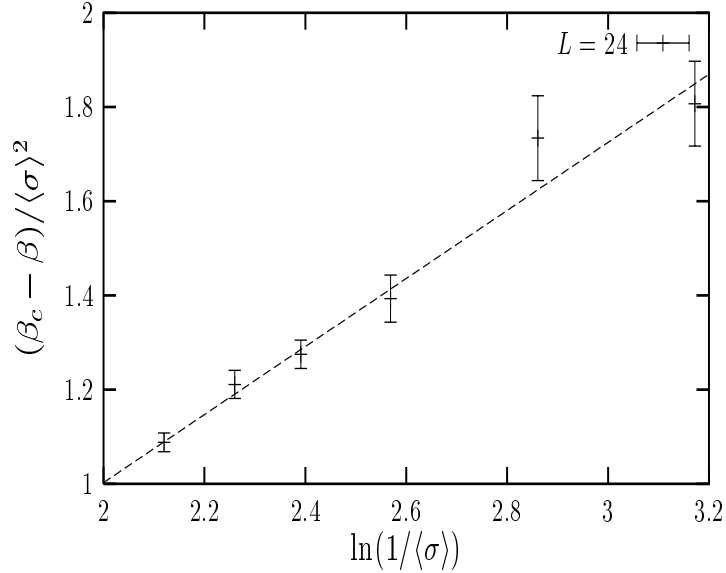}
}
\caption{$(\beta_c-\beta)/\langle \sigma \rangle^2$ vs $\ln (1/\langle \sigma \rangle)$ 
at Four Fermi coupling $\lambda=2.0$.}
\label{fig:sig24}
\end{figure}

The raw data in Fig.~\ref{fig:pqed24bs2sig} is fit with a simple power, 
$\langle \sigma \rangle = A(\beta_c-\beta)^{\beta_{mag}}$,
and the parameters $A = 2.8(9)$, $\beta_c = 0.952(1)$, $\beta_{mag} = 0.77(10)$, were determined with
$\chi^2/DOF = 1.18$. We again see that the best power law predicts a critical
index $\beta_{mag}$ higher than that of pure mean field theory, as in reference \cite{KC}. However, the data
is also well fit with the hypothesis of triviality. Fits to the form
$\beta_c-\beta=A \langle \sigma \rangle^2 (\ln (1/\langle \sigma \rangle) +B)$
gave the parameters $A = 0.21(10)$, $\beta_c = 0.951(1)$ and $B = -0.061(1)$
with a fine quality of fit
$\chi^2/DOF = 0.78$. Following our presentation above, we plot $(\beta_c-\beta)/\langle \sigma \rangle^2$ against
$\ln (1/\langle \sigma \rangle)$ in Fig.~\ref{fig:sig24}, which shows the importance of the logarithms and
the deviation from pure mean field theory.
As in Sec.~4 we find that the log-improved triviality fit is preferred to the power law form.

\section{Results near the critical coupling}
\subsection{Background on Finite Size Scaling}

In order to study the critical behavior of our theory arbitrarily close to the critical coupling we used
FSS methods. FSS techniques first developed by Fisher \cite{Fisher} 
are important tools used in the determinations of critical exponents
near second order phase transitions.
The critical behavior of a system in the thermodynamic limit may be extracted from
the properties of finite size systems by examining the size dependence of the singular part
of the free energy density.
According to FSS theory, for dimensionality $d$ less than the upper critical dimension $d_c$,
 the singular part of the free energy is described
phenomenologically by a universal scaling form,
\begin{equation}
F_s(t,m,L)=L^{-d} \mathcal{F}(tL^{1/\nu}, mL^{(\beta_{mag}+\gamma)/\nu}),
\label{eq:free_energy}
\end{equation}
where $m$ is the fermion bare mass and $t \equiv (\beta_c - \beta)$.
The critical exponents $\nu$, $\beta_{mag}$ and $\gamma$ are all the thermodynamic values
for the infinite system.
Scaling forms for various thermodynamic quantities can be obtained from
appropriate derivatives of the free energy density.
On a finite volume and with the fermion bare mass set to zero, the direction of
symmetry breaking changes over the course of the run so the chiral condensate
averages to zero over the ensemble. Another option is to introduce an effective
order parameter $\Sigma \equiv \langle |\sigma| \rangle$, which in the thermodynamic limit
is equal to the true order parameter $\langle \sigma \rangle$.
The FSS scaling form for $\Sigma$ determined from Eq.~\ref{eq:free_energy} is
\begin{equation}
\Sigma  =  L^{-\beta_{mag}/\nu}f_{\sigma}(tL^{1/\nu}).
\label{eq:magn1}
\end{equation}

A standard method to measure $\beta_c$ for a second order transition is to compute
the Binder cumulant \cite{Binder} for various system sizes. 
On sufficiently large lattices where subleading corrections from the finite lattice size $L$
are negligible, the Binder cumulant $U_B(\beta,L)$, defined by
\begin{equation}
U_B \equiv 1-\frac{1}{3}\frac{\langle|\sigma|^4\rangle}{\langle|\sigma|^2\rangle^2},
\end{equation}
is given by $U_B=f_B L(tL^{1/\nu})$ and, therefore, at $\beta_c$ it becomes independent of $L$.

Another quantity of interest is the susceptibility $\chi$ which, in the static limit of the
fluctuation-dissipation theorem, is 
\begin{equation}
\chi = \lim_{L \to \infty} V[\langle \sigma^2 \rangle - \langle \sigma \rangle^2],
\end{equation}
where $V$ is the lattice volume.
For finite systems this expression leads to the following finite-lattice estimates for $\chi$:
\begin{equation}
\chi_1=V\langle \sigma^2 \rangle  \: \: \: \: \beta  >  \beta_c,  \\
\label{eq:chi1}
\end{equation}
\begin{equation}
\chi_c = V[\langle \sigma^2 \rangle - \langle |\sigma| \rangle^2]   \: \: \: \: \beta < \beta_c,
\label{eq:chi_c}
\end{equation}
where the subscript $c$ stands for ``connected.'' Both relations should scale at criticality like $L^{\gamma/\nu}$.
Furthermore, the maxima of $\chi_c$ in the scaling region should also obey
$\chi_c^{\rm peak} \sim L^{\gamma/\nu}$.

Furthermore, logarithmic derivatives of $\langle |\sigma|^n \rangle$ can give additional estimates for $\nu$.
It is easy to show that  
\begin{equation}
D_n \equiv \frac{\partial}{\partial \beta} {\rm ln}\langle |\sigma|^n \rangle  =  
\left[ \frac{\langle |\sigma|^n P \rangle}
{\langle |\sigma|^n \rangle} - \langle P \rangle \right],
\label{eq:derlog}
\end{equation}
where $P$ is the plaquette, has a scaling relation 
\begin{equation}
D_n = L^{1/\nu}f_{D_n}(tL^{1/\nu}).
\end{equation}

Other related quantities useful in determining the critical exponent $\nu$ can be defined
from logarithms of derivatives of $\langle \sigma^n \rangle$ \cite{Chen}. In our analysis we will consider
\begin{equation}
Q \equiv 2[\sigma^2] - [\sigma^4],
\end{equation}
where
\begin{equation}
[\sigma^n] \equiv \ln \frac{\partial \langle \sigma^n \rangle}{\partial \beta}.
\end{equation}
One can easily show that
\begin{equation}
Q \simeq \frac{1}{\nu} \ln L + \mathcal{Q}(tL^{1/\nu}).
\label{eq:Vcum}
\end{equation}

The above FSS relations rely on the traditional FSS hypothesis that in the vicinity of the critical coupling 
the behavior of the system is determined by the scaled variable $L/\xi$, where $\xi$ is the correlation 
length. The standard FSS hypothesis fails for $d \geq d_c$. A modified hypothesis for $O(N)$-symmetric
$\Phi^4$ theories in 
four dimensions was proposed in \cite{Kenna}, where it was shown that in the vicinity of the critical 
coupling the actual length of the finite size system is replaced by its correlation length
$\xi_L(0) \propto L({\rm ln}L)^\frac{1}{4}$, independent of $N$. 
However, as shown in \cite{Looking} and demonstrated numerically in earlier sections of this paper, the
logarithmic triviality in fermionic field theories such as the NJL model and QED is manifested in a different
way from the triviality in purely bosonic theories. The logarithmic corrections in the scaling relations
of QED are expected to be in the denominator of the scaling functions, whereas in $\Phi^4$ theory they are in 
the numerator.
The same is expected in the FSS relations of various thermodynamic quantities.
By generalizing the Privman-Fisher ansatz for the scaling relation of the singular part 
of the free energy, the scaling function for $F_s(t,m_0,L)$ 
becomes
\begin{equation}
F_s(t,m_0,L)=L^{-d}\mathcal{F}(tL^2{\rm ln}^pL, m_0L^3 {\rm ln}^qL).
\end{equation}
Consequently, the log-improved FSS relations for the effective order parameter and the susceptibility 
obtained from appropriate derivatives at zero fermion mass are, 
\begin{equation}
\Sigma  =  L^{-1}{\rm ln}^qL f_{\sigma}(tL^2{\rm ln}^pL),
\label{eq:log1}
\end{equation}
\begin{equation}
\chi  =  L^{2}{\rm ln}^{2q}L f_{\chi}(tL^2{\rm ln}^pL).
\label{eq:log2}
\end{equation}

In the next sections we will present our attempts to extract the critical exponents and to check
consistency with log-improved scaling. We studied the FSS behavior of several observables in 
order to compare different results and reach conclusions. 

\subsection{Analysis of the Binder cumulant}
In the vicinity of $\beta_c$ we can expand the Binder cumulant and find
\begin{equation}
U_B(\beta,L) \simeq U_* + U_1(\beta_c-\beta)L^{1/\nu}. 
\end{equation}
By fitting this relation to data in the range $1.2875 \leq \beta \leq 1.375$ and lattice sizes
$L=8,..,16$ we found $\nu=0.49(10)$, $\beta_c=1.356(7)$ and $U_*=0.153(26)$ with
$\chi^2/DOF=1.6$. The measured value of the critical exponent $\nu$ is consistent with
the mean field prediction $\nu=1/2$, although the error is relatively large.
More measurements of $\nu$ with better precision are presented below. The location of the critical
point $\beta_c=1.356(7)$ refines the estimates found from the broken symmetry fits given earlier. Since
those studies were done far from $\beta_c$, a small discrepancy is not surprising.

\begin{figure}[htb]
\bigskip\bigskip
\begin{center}
\epsfig{file=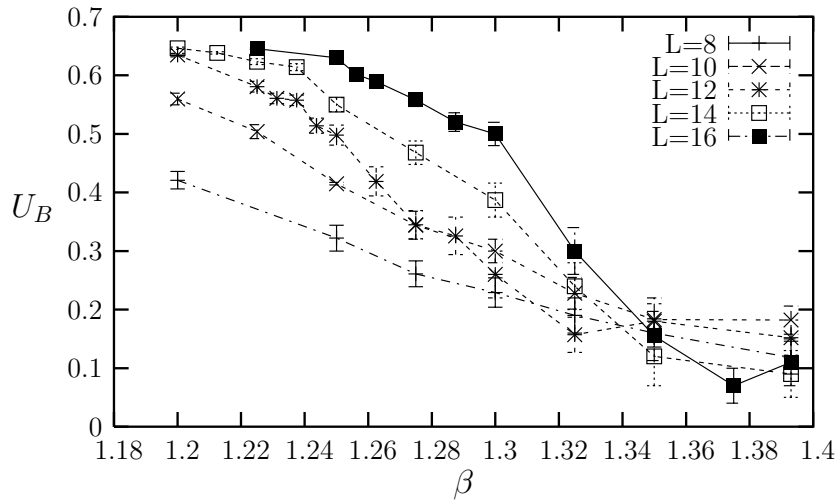, width=11cm}
\end{center}
\caption{Binder cumulant versus coupling for different lattice sizes.}
\label{fig:fig1}
\end{figure}

\subsection{Analysis of effective order parameter}
\begin{figure}[t]
\bigskip\bigskip
\begin{center}
\epsfig{file=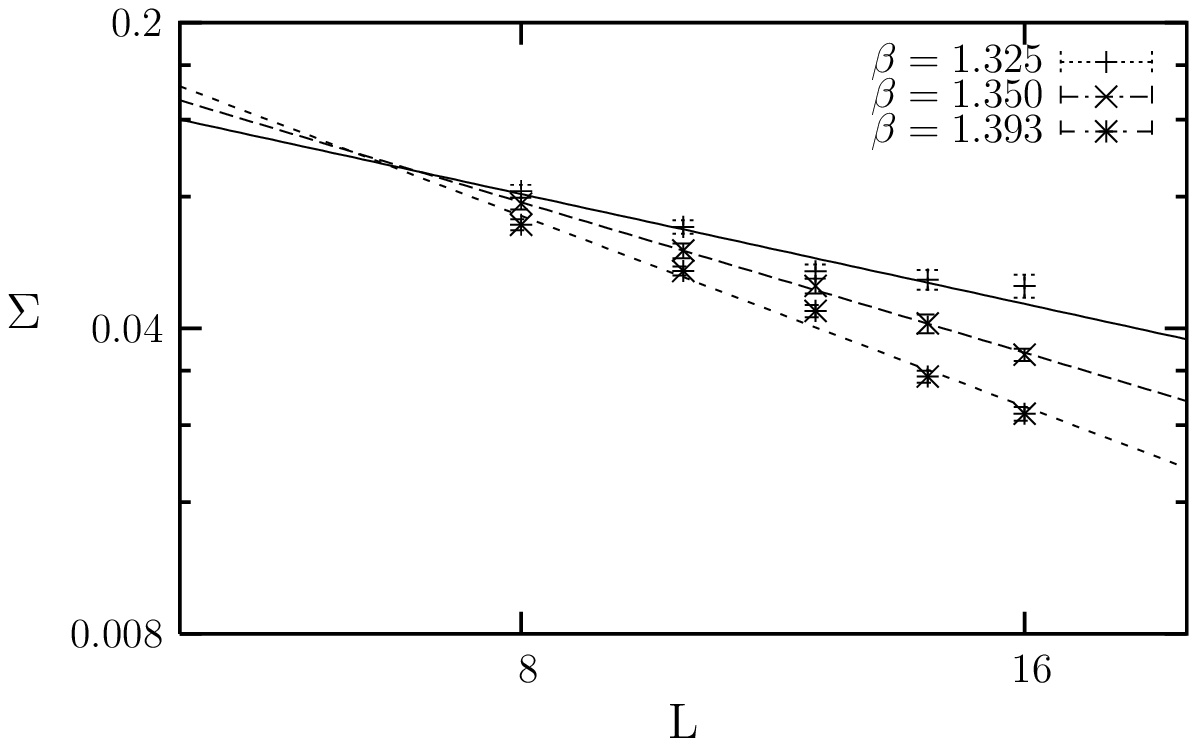, width=10cm}
\end{center}
\caption{$\Sigma$ vs. $L$ for different values of $\beta$.}
\label{fig:fig2}
\end{figure}

In this section we discuss the FSS analysis for the effective order parameter $\Sigma$. We fit our data
to $\Sigma_{\beta_c} = aL^{-\beta_{mag}/\nu}$ for three values of $\beta=1.325, 1.350$ and $1.393$, 
which are close to the value $\beta_c=1.356$ extracted from the analysis of $U_B(L,\beta)$.
The results are presented in Table \ref{tab:t1} and plotted in Fig.~\ref{fig:fig2}.
We conclude that $\beta_c$ is close to $1.350$ and the ratio $\beta_{mag}/\nu=1.14(6)$ at this value
of $\beta$ is also close to the mean field result $\beta_{mag}/\nu=1$.

\begin{figure}[t]
\bigskip\bigskip
\begin{center}
\epsfig{file=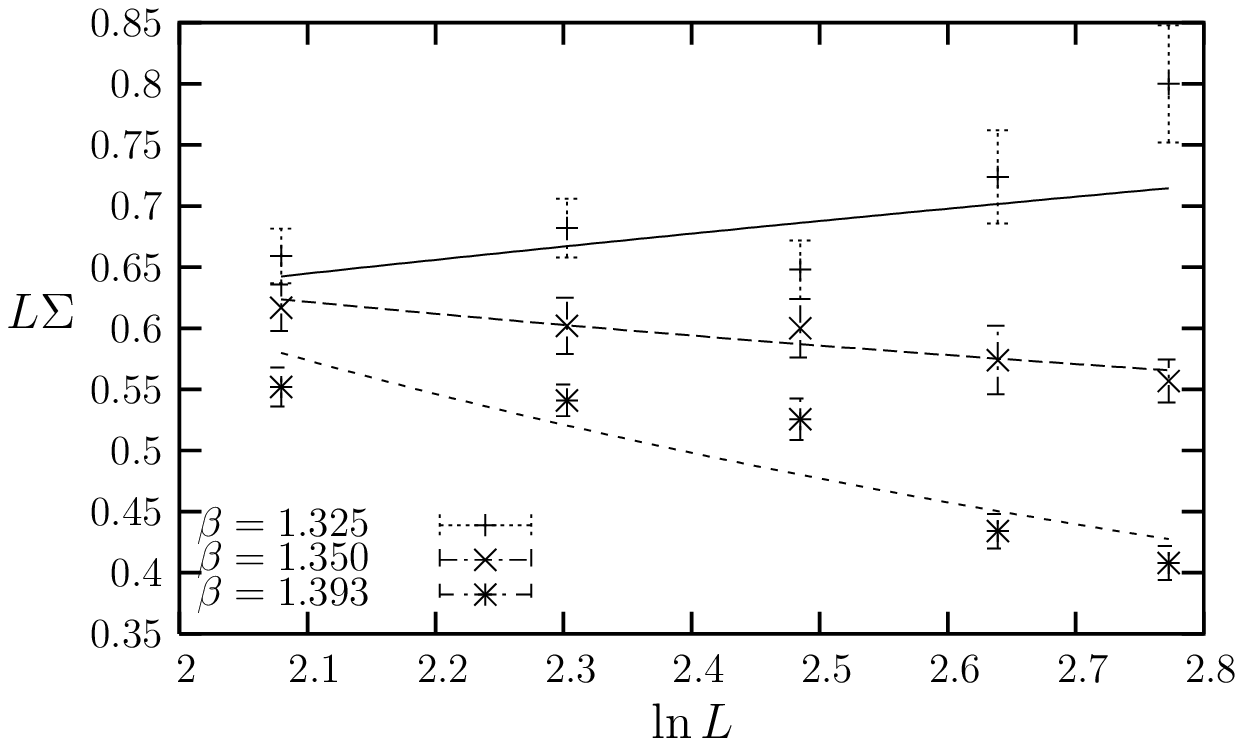, width=11cm}
\end{center}
\caption{Fits to $L \Sigma = a \ln^p L$ for different values of $\beta$ near the transition.}
\label{fig:fig3}
\end{figure}

We also fit all the data in the vicinity of the transition to a single scaling function obtained
from the Taylor expansion of Eq.~9, up to a linear term,
\begin{equation}
\Sigma \simeq [c_1 + c_2(\beta_c-\beta)L^{1/\nu}]L^{\beta_{mag}/\nu}.
\end{equation}
After fixing $\beta_c=1.356$ we  obtained the values of  $\beta_{mag}/\nu$ and $\nu$. 
When all the three values of the coupling ($\beta=1.325, 1.350$ and $1.393$) for $L=8,...,16$,
and an extra coupling $\beta=1.375$ for $L=16$,
are included in the fit
we get $\beta_{mag}/\nu=1.13(4)$ and $\nu=0.48(8)$ with $\chi^2/DOF=1.4$.
For the same data set and for fixed $\beta_c=1.330$ (which is the value of the critical coupling obtained from the 
broken phase analysis) we get $\beta_{mag}/\nu=0.95(5)$ and $\nu=0.56(8)$ with $\chi^2/DOF=1.3$. 
  
In order to check whether our results are consistent with log-improved mean field scaling, 
we fit $\Sigma$ to Eq.~\ref{eq:log1}.
The results are summarized in Table \ref{tab:t2} and plotted in Fig.~\ref{fig:fig3}. The best fit is at
$\beta=1.350$ with $p=-0.34(14)$.
The negative sign of $p$ is consistent with the scenario of log-triviality in 
fermionic field theories \cite{kocic}.

\subsection{Analysis of susceptibility}
\begin{figure}[]
\bigskip\bigskip
\begin{center}
\epsfig{file=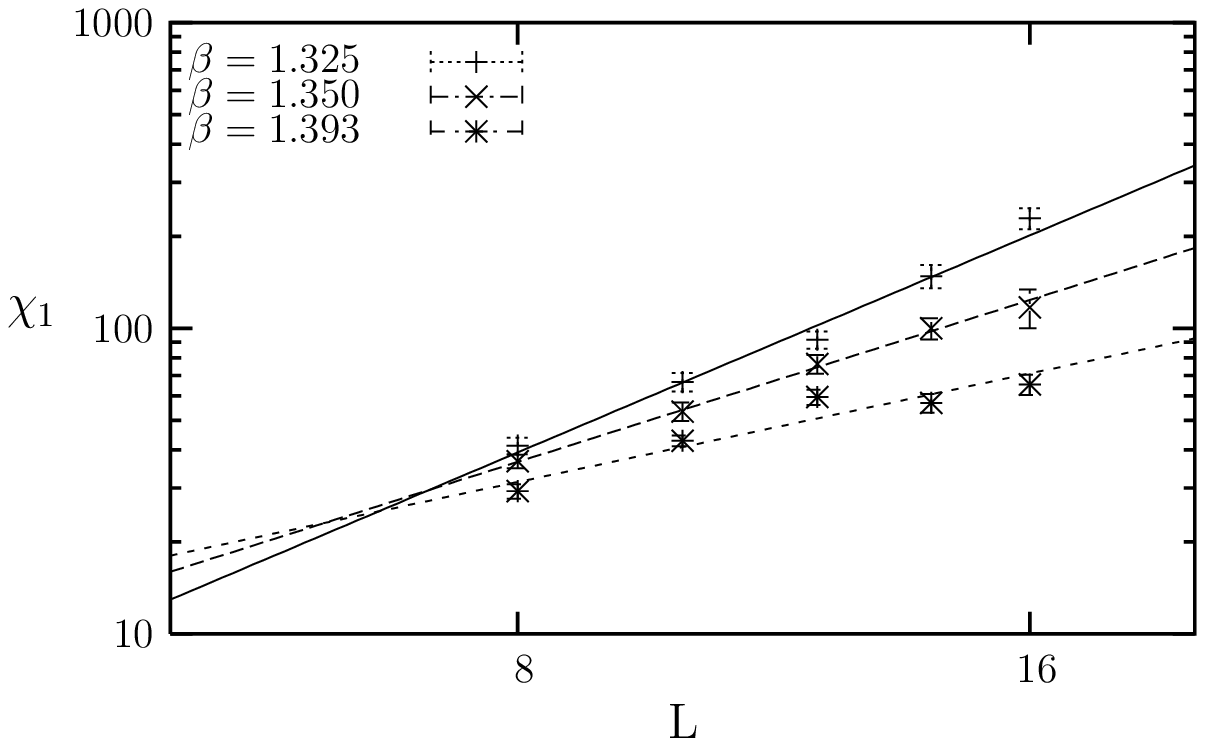, width=10cm}
\end{center}
\caption{Susceptibility $\chi_1$ vs. $L$ for different values of $\beta$.}
\label{fig:susc1}
\end{figure}

First, we fit $\chi_1$ (Eq.~\ref{eq:chi1}) as a function of $L$ at different values of  
$\beta =1.325, 1.350$ and $1.393$.
The results are displayed in Fig.~\ref{fig:susc1} and Table \ref{tab:t3}. It is clear from these results that the
critical coupling is close to $\beta=1.350$ in agreement with analyses presented in previous paragraphs.
The measured value of $\gamma/\nu=1.76(14)$ is close
to its mean field value of $2$. We also fit $\chi_1$ in the vicinity of the transition
to the linear expansion of $\chi(L,\beta)$ 
around $\beta_c$.
\begin{equation}
\chi_1(L,\beta) \simeq [c_1 + c_2(\beta_c-\beta)L^{1/\nu}]L^{\gamma/\nu}.
\end{equation}

After fixing the critical coupling to the value extracted from the broken phase analysis ($\beta_c=1.330$)
we get $\nu=0.66(9)$ and $\gamma/\nu=2.15(9)$ with $\chi/{\rm DOF=1.3}$, whereas after fixing $\beta_c$ to
the value extracted from the $U_B$ analysis we get $\nu=0.51(7)$ and $\gamma/\nu=1.87(8)$ with the 
same $\chi^2/{\rm DOF}$
as before. Furthemore, we fit the data to $\chi_1 \simeq a L^2 ({\rm \ln}^{2p}L)$ in order to check whether
the data are consistent with log-improved mean field scaling.
The results 
displayed in Table~\ref{tab:t5} provide good evidence that the log-improved scaling relation describes the data well
and that $\beta_c \simeq 1.350$. The measured value $p=-0.28(16)$ is compatible
with the result extracted from log-improved fits of $\Sigma$ discussed in the previous section.
\begin{figure}[]
\bigskip\bigskip
\begin{center}
\epsfig{file=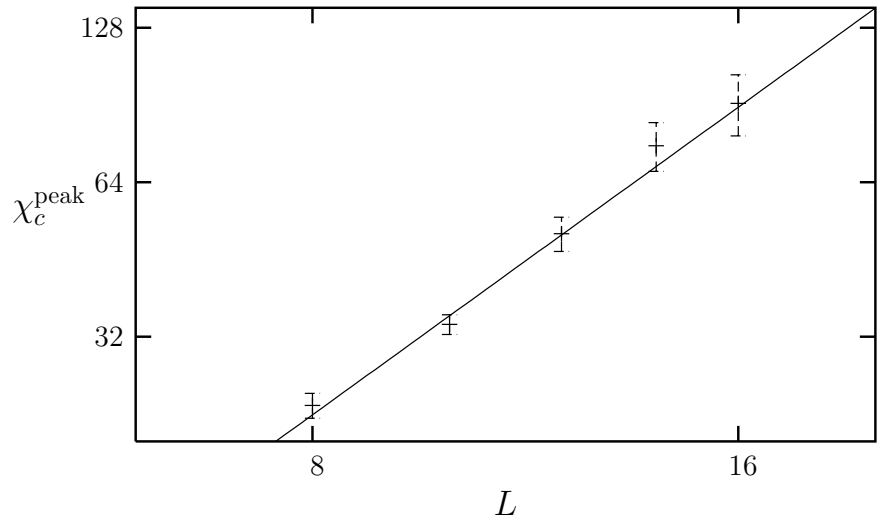, width=10cm}
\end{center}
\caption{Peaks of connected susceptibility $\chi_c$ vs. $L$.}
\label{fig:chi_c}
\end{figure}

We repeated the above analysis for the connected susceptibility $\chi_c$. The results presented 
in Table~\ref{tab:t5} indicate clearly
that $\beta_c$ is close to $\beta \simeq 1.350$ with the value $\gamma/\nu=1.70(11)$ close to the mean field result. 
The results of fits to the log-improved FSS relation are summarized in Table~\ref{tab:t6}. Again the results
provide significant
evidence that $\beta_c$ is close to $1.350$ and $p=-0.36(13)$ which is consistent with previous measurements of $p$.
We also fitted the peaks of $\chi_c$ to $\chi_c^{\rm peak} \sim L^{\gamma/\nu}$ and got $\gamma/\nu=1.99(16)$
in good agreement with the mean field prediction. The data and the fitting function are shown in Fig.~\ref{fig:chi_c}.

\subsection{Analysis of $D_j$  and $Q$}
To make a further check of our results for the exponent $\nu$, we studied the finite size scaling properties of
logarithmic derivatives of $\langle |\sigma|^n \rangle$ defined in Eq.~\ref{eq:derlog}.
We fit $D_j \sim L^{1/\nu}$ for $j=1,2,$ and $3$
at different values of the gauge coupling and for $L=8,...,14$.  We note that the data generated on
$16^4$ lattices were noisy and therefore could not be included in the fits.
The results for the exponent $\nu$ and the quality of each fit are shown
in Tables \ref{tab:t7},\ref{tab:t8},\ref{tab:t9}. ithe values of $\nu$ are in very good 
agreement with the mean field prediction $\nu=0.5$.
Although for these values of gauge coupling $\nu$ do not have a significant dependence on $\beta$, 
the qualities of the fits indicate
that $\beta_c \simeq 1.350$ which
is in agreement with results presented in previous paragraphs. Our attempts to fit the data to log-improved FSS
mean field scaling laws $D_j \sim L^2 \ln^p L$ did not give any signal for $p$. 
\begin{figure}[htb]
\bigskip\bigskip
\begin{center}
\epsfig{file=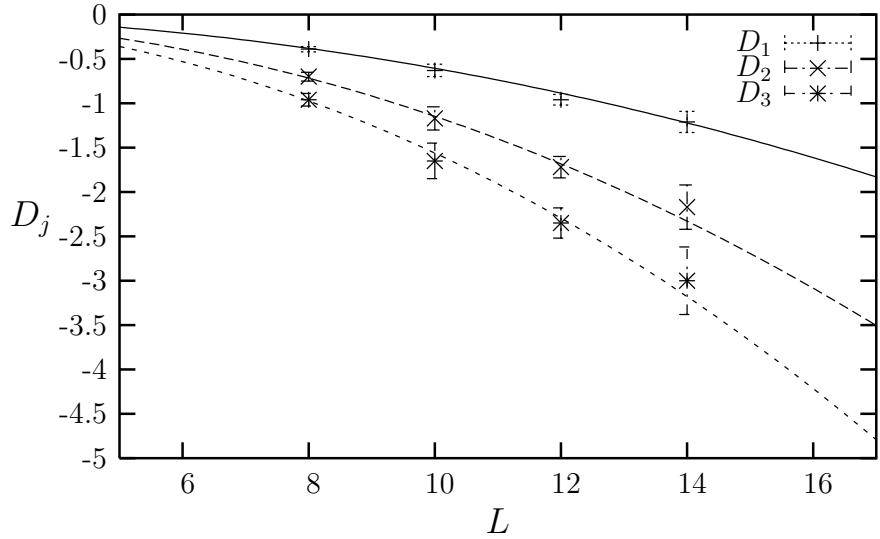, width=11cm}
\end{center}
\caption{$D_1,D_2$ and $D_3$ vs. $L$ at $\beta=1.350$.}
\label{fig:ferm_disp2}
\end{figure}

Furthermore, we fit all the data ($\beta =1.325,1.350,1.393$) to the linear expansion of the FSS relation
\begin{equation}
D_j(\beta,L) \simeq c_{1j}L^{1/\nu} + c_{2j}(\beta_c-\beta)L^{2/\nu},
\label{eq:Dj}
\end{equation}
with fixed $\beta_c=1.356$. 
The results are summarized in Table \ref{tab:t10} and are in good agreement with
the mean field result although the quality of each fit
is not as good as before.

Finally, a fit to the linear expansion of the observable $Q$ (defined in Eq.~\ref{eq:Vcum})
\begin{equation}
Q \approx \frac{1}{\nu}\ln L + c_1 + c_2 (\beta_c-\beta)L^{1/\nu}
\end{equation}
for $\beta=1.300,...,1.393$,  $L=8,...,14$ and fixed $\beta_c=1.356$ gives $\nu=0.52(3)$
and $\chi^2/DOF =1.5$, in good agreement with all the measurements for $\nu$ presented in
previous paragraphs.

\section{Conclusions}

We have presented evidence for the logarithmic triviality of the chiral transition line
in compact QED with four species of fermions. A weak four-fermi interaction was employed
in the action so that massless fermions could be simulated directly, thus avoiding the
need to extrapolate raw data to the chiral limit. $\chi$QED allowed us to use the simplest
single variable finite size scaling fits to the data. Since
the four-fermi interaction is irrelevant in four dimensions and since the
full model $\chi$QED is found to be logarithmically trivial, this study constitutes 
strong evidence that the continuum limit of the standard
compact lattice QED model is also logarithmically trivial.

Finite size scaling proved to be an effective approach
to deciding the physics issues inherent in these models.
The FSS analyses provide strong evidence that the critical exponents are the mean field theory ones.
Especially in the case of $\nu$ the analyses of various observables show in a
consistent manner that $\nu$  is very close to
0.5. The broken phase data at different values of the four-fermi coupling also
favor fermionic log-improved mean field scaling
over the interacting field theory scenario.

\section*{Acknowledgments}

JBK was partially supported by NSF under grant NSF-PHY0304252
The simulations were done at PSC, NPACI, NCSA and NERSC. Special thanks go to the
staff of PSC, especially David O'Neal, for help with code optimization and
parallel processing. We also thank NERSC for their longterm, steady support.

\begin{table}[ht]
\centering
\caption{Results from fits to $\Sigma = a L^{-\beta_{mag}/\nu}$.}
\medskip
\label{tab:t1}
\begin{tabular}{|c|c|c|c|}
\hline
$\beta$ & $\beta_{mag}/\nu$ & $a$ &  $\chi^2/{\rm DOF}$ \\
\hline \hline
1.325  & 0.84(8) &  0.46(9)   &  2.1  \\
1.350  & 1.14(6) &  0.84(12)  & 0.16 \\
1.393  & 1.45(5) &  1.48(20)  &  4.6  \\
\hline
\end{tabular}
\end{table}

\begin{table}[ht]
\centering
\caption{Results from fits to $\Sigma = a L^{-1}\ln^p L$.}
\medskip
\label{tab:t2}
\begin{tabular}{|c|c|c|c|}
\hline
$\beta$ & $a$    & $p$        &  $\chi^2/{\rm DOF}$ \\
\hline \hline
1.325  & 0.49(8) &  0.37(19)   &  2.2  \\
1.350  & 0.80(10) &  -0.34(14)  & 0.20 \\
1.393  & 1.26(14) &  -1.06(13)  &  5.4 \\
\hline
\end{tabular}
\end{table}

\begin{table}[]
\centering
\caption{Fits for $\chi_1 = a L^{\gamma/\nu}$.}
\medskip
\label{tab:t3}
\begin{tabular}{|c|c|c|c|}
\hline
$\beta$ & $a$    & $\gamma/\nu$  &  $\chi^2/{\rm DOF}$ \\
\hline \hline
1.325  & 0.29(10) &  2.36(14)  &  2.0  \\
1.350  & 0.94(30) &  1.76(14)  & 0.12 \\
1.393  & 2.7(7)   &  1.18(11)  &  4.0 \\
\hline
\end{tabular}
\end{table}

\begin{table}[]
\centering
\caption{Results from fits to $\chi_1 = a L^2 (\ln^{2q} L)$.}
\medskip
\label{tab:t4}
\begin{tabular}{|c|c|c|c|}
\hline
$\beta$ & $a$    & $p$  &  $\chi^2/{\rm DOF}$ \\
\hline \hline
1.325  & 0.33(10) &  0.41(17)  &  2.2  \\
1.350  & 0.86(23)  &  -0.28(16)  & 0.13 \\
1.393  &  2.06(45)  &  -0.97(13) & 4.7  \\
\hline
\end{tabular}
\end{table}

\begin{table}[]
\centering
\caption{Results from fits to $\chi_c = a L^{\gamma/\nu}$.}
\medskip
\label{tab:t5}
\begin{tabular}{|c|c|c|c|}
\hline
$\beta$ & $a$    & $\gamma/\nu$  &  $\chi^2/{\rm DOF}$ \\
\hline \hline
1.325  & 0.056(15) &  2.51(11)  &  31.3  \\
1.350  & 0.36(9)  &  1.70(11)  & 0.9 \\
1.393  & 0.80(18)   &  1.23(9)  &  4.5 \\
\hline
\end{tabular}
\end{table}

\begin{table}[t]
\centering
\caption{Results from fits to $\chi_c = a L^2(\ln^{2p}L)$.}
\medskip
\label{tab:t6}
\begin{tabular}{|c|c|c|c|}
\hline
$\beta$ & $a$    & $p$  &  $\chi^2/{\rm DOF}$ \\
\hline \hline
1.325  & 0.11(3) &  0.30(17)  &  45.6  \\
1.350  & 0.33(7)  &  -0.36(13)  & 0.9 \\
1.393  & 0.62(12)   &  -0.91(11)  &  5.1 \\
\hline
\end{tabular}
\end{table}

\begin{table}[]
\centering
\caption{Results of fits to $D_1 \sim L^{1/\nu}$ }
\medskip
\label{tab:t7}
\begin{tabular}{|c|c|c|}
\hline
$\beta$ & $\nu$   &  $\chi^2/{\rm DOF}$ \\
\hline \hline
1.325  &  0.46(4)  &  1.1   \\
1.350  &  0.48(4)  &  0.4  \\
1.393  &  0.47(6)  &  3.3  \\
\hline
\end{tabular}
\end{table}
\begin{table}[]
\centering
\caption{Results of fits to $D_2 \sim L^{1/\nu}$ }
\medskip
\label{tab:t8}
\begin{tabular}{|c|c|c|}
\hline
$\beta$ & $\nu$   &  $\chi^2/{\rm DOF}$ \\
\hline \hline
1.325  &  0.47(4)  &  1.0   \\
1.350  &  0.47(5)  &  0.3  \\
1.393  &  0.48(7)  &  2.8  \\
\hline
\end{tabular}
\end{table}

\begin{table}[]
\centering
\caption{Results of fits to $D_3 \sim L^{1/\nu}$ }
\medskip
\label{tab:t9}
\begin{tabular}{|c|c|c|}
\hline
$\beta$ & $\nu$   &  $\chi^2/{\rm DOF}$ \\
\hline \hline
1.325  &  0.49(4)  &  1.0   \\
1.350  &  0.47(5)  &  0.3  \\
1.393  &  0.51(7)  &  2.3  \\
\hline
\end{tabular}
\end{table}

\begin{table}[]
\centering
\caption{Results of fits to eq.~\ref{eq:Dj}. }
\medskip
\label{tab:t10}
\begin{tabular}{|c|c|c|}
\hline
$ $     &    $\nu$   &  $\chi^2/{\rm DOF}$ \\
\hline \hline
$D_1$   &  0.49(3)  &  2.5   \\
$D_2$  &  0.51(3)   &  2.0  \\
$D_3$  &  0.51(3)   &  1.9  \\
\hline
\end{tabular}
\end{table}

\end{document}

%% file: pqedz2.pstex_t
\begin{picture}(0,0)%
\includegraphics{pqedz2.pstex}%
\end{picture}%
\setlength{\unitlength}{4144sp}%
\begingroup\makeatletter\ifx\SetFigFont\undefined%
\gdef\SetFigFont#1#2#3#4#5{%
  \reset@font\fontsize{#1}{#2pt}%
  \fontfamily{#3}\fontseries{#4}\fontshape{#5}%
  \selectfont}%
\fi\endgroup%
\begin{picture}(9067,6145)(361,-5744)
\put(3286,-5551){\makebox(0,0)[lb]{\smash{\SetFigFont{17}{20.4}{\familydefault}{\mddefault}{\updefault}{\color[rgb]{0,0,0}$0.25$}%
}}}
\put(4456,-5686){\makebox(0,0)[lb]{\smash{\SetFigFont{17}{20.4}{\familydefault}{\mddefault}{\updefault}{\color[rgb]{0,0,0}$G=1/\lambda$}%
}}}
\put(5896,-691){\makebox(0,0)[lb]{\smash{\SetFigFont{17}{20.4}{\familydefault}{\mddefault}{\updefault}{\color[rgb]{0,0,0}$<\bar{\psi}\psi> \not= 0$   $M = 0$}%
}}}
\put(1351,-691){\makebox(0,0)[lb]{\smash{\SetFigFont{17}{20.4}{\familydefault}{\mddefault}{\updefault}{\color[rgb]{0,0,0}$<\bar{\psi}\psi> = 0$   $M = 0$}%
}}}
\put(361,-511){\makebox(0,0)[lb]{\smash{\SetFigFont{17}{20.4}{\familydefault}{\mddefault}{\updefault}{\color[rgb]{0,0,0}$\beta$}%
}}}
\put(451,-2491){\makebox(0,0)[lb]{\smash{\SetFigFont{17}{20.4}{\familydefault}{\mddefault}{\updefault}{\color[rgb]{0,0,0}$1.0$}%
}}}
\put(2926,-3976){\makebox(0,0)[lb]{\smash{\SetFigFont{17}{20.4}{\familydefault}{\mddefault}{\updefault}{\color[rgb]{0,0,0}$<\bar{\psi}\psi> \not= 0$   $M \not= 0$}%
}}}
\end{picture}